\documentclass[runningheads]{llncs}
\usepackage[T1]{fontenc}
\usepackage{graphicx,amssymb,amsmath}

\newcommand\bR{\mathbb{R}}
\newcommand\bE{\mathbb{E}}

\frontmatter

\begin{document}

\mainmatter              

\title{Value of Information in the Mean-Square Case and its Application to the Analysis of Financial Time-Series Forecast}
\titlerunning{VoI in the Mean-Square Case and Analysis of Time-Series Forecast}

\author{Roman V. Belavkin\inst{1}\orcidID{0000-0002-2356-1447} \and
  Panos Pardalos\inst{2}\orcidID{0000-0001-9623-8053} \and
  Jose Principe\inst{3}\orcidID{0000-0002-3449-3531}}

\authorrunning{R. V. Belavkin et al.}

\institute{Department of Computer Science, Middlesex University, London, NW4 4BT, UK \email{r.belavkin@mdx.ac.uk}\and
Department of Industrial and Systems Engineering, University of Florida, P.O. Box 116595, Gainesville, FL 32611-6595, USA\\
\email{pardalos@ufl.edu} \and
Department of Electrical \& Computer Engineering, University of Florida, P.O. Box 116130, Gainesville, FL 32611-6130, USA\\
\email{principe@cnel.ufl.edu}}

\maketitle              

\begin{abstract}
The advances and development of various machine learning techniques has lead to practical solutions in various areas of science, engineering, medicine and finance.  The great choice of algorithms, their implementations and libraries has resulted in another challenge of selecting the right algorithm and tuning their parameters in order to achieve optimal or satisfactory performance in specific applications.  Here we show how the value of information ($V(I)$) can be used in this task to guide the algorithm choice and parameter tuning process.  After estimating the amount of Shannon's mutual information between the predictor and response variables, $V(I)$ can define theoretical upper bound of performance of any algorithm.  The inverse function $I(V)$ defines the lower frontier of the minimum amount of information required to achieve the desired performance.  In this paper, we illustrate the value of information for the mean-square error minimization and apply it to forecasts of cryptocurrency log-returns.
\end{abstract}

\section{Introduction}\label{sec:intro}

The value of information $V(I)$ is the maximum gain in performance one can achieve due to receiving the amount $I$ of information (mathematical meaning of `performance' and `information' will be clarified later).  This concept was discussed in various settings in the literature, but the main advances of the theory behind it were made by Ruslan Stratonovich and his colleagues in the 1960s \cite{Stratonovich65,Stratonovich-Grishanin66,Stratonovich66:_value_automata,Grishanin-Stratonovich66,Stratonovich67:_inf_dyn,Stratonovich-Grishanin68}.  Inspired by Shannon's rate-distortion theory \cite{Shannon48}, Stratonovich first extended the ideas to more general class of Bayesian systems and various types of information.  He then used original techniques and some methods of statistical physics to derive very deep results on asymptotic equivalence of the value functions for different types of information.  Stratonovich and his colleagues also studied the value of information in different settings, from the simplest Boolean and Gaussian systems to stochastic processes in continuous time.  Many of these examples are covered in the classical monograph \cite{Stratonovich75:_inf}, which has recently been published in English \cite{Stratonovich20:_inf}.

Recent advances of intelligent and learning systems combined with exponential growth of the size and dimensionality of datasets facilitated by the growth in computer performance has prompted a new interest in the value of information theory and its applications.  Some results of the theory have facilitated better understanding of the role of randomization in machine learning algorithms \cite{Belavkin09:_dyninf,Belavkin09:_adprl09,Belavkin11:_optim}.  For example, the value of information was used to derive optimal control functions of mutation rates in genetic algorithms \cite{Belavkin11:_itw11,Belavkin11:_dyninf,Belavkin_etal16:_jomb}.  It was shown also that the value of information theory is closely related to optimal transport \cite{Belavkin18:_igaia} and can have unexpected applications in explaining some decision-making paradoxes in behavioural economics \cite{Belavkin14:_risk}.

The purpose of this paper is to demonstrate how the value of information can be used to evaluate the performance and tune parameters of different data-driven models with a specific focus on the mean-square error criterion.  In the next section, we briefly overview the VoI theory for the case of translation invariant objective functions, such as the mean-square deviation.  We derive a simple expression for the smallest root-mean-square error (RMSE) as a function of Shannon's mutual information between the predictor and response variables.  This function is then used in Section~\ref{sec:rmse} as performance frontier for several models attempting to forecast daily log-returns of some cryptocurrencies.  We conclude by the discussion of these results, the importance of correct estimation of the amount of information in data as well as the choice of objective functions to evaluate the models.

\section{Value of information for translation invariant objective functions}\label{sec:voi-theory}

Let us review some of the main ideas of the value of information theory in the context of optimal estimation, although the context of optimal control is also relevant.  Let $(\Omega,P,\mathcal{A})$ be a probability space, and let $x\in X$ be a random variable (i.e. a measurable function $x=x(\omega)$ on a probability space, and $P(X)=P\{\omega:x(\omega)\in X\}$ is the corresponding push-forward measure).  Consider the problem of finding an element $y\in Y$ maximizing the expected value of \emph{utility} function $u:X\times Y\to\bR$.  Let us denote the corresponding optimal value as follows:
\[
  U(0):=\sup_{y\in Y}\bE_{P(x)}\{u(x,y)\}
\]
where zero in $U(0)$ designates the fact that no information about specific value of $x\in X$ is given, only the prior distribution $P(x)$.  At the other extreme, full information entails that there is an invertible function $z=f(x)$ such that $x\in X$ is determined uniquely $x=f^{-1}(z)$ by the `message' $z\in Z$.  The corresponding optimal value is
\[
  U(\infty):=\bE_{P(x)}\{\sup_{y(z)}u(x,y(z))\}
\]
where optimization is over all mappings $y(z)$ (i.e. $y:Z\to Y$).  In the context of estimation, variable $x$ is the \emph{response} (i.e. the variable of interest), and $z$ is the \emph{predictor}.  The mapping $y(z)$ represents a model with output $y\in Y$.

Let us denote by $U(I)$ the intermediate values in the interval $[U(0),U(\infty)]$ for all information amounts $I\in[0,\infty]$.  The value of information is then defined as the following difference \cite{Stratonovich20:_inf}:
\[
  V(I):=U(I)-U(0)
\]

There are, however, different ways in which information amount $I$ and the quantity $U(I)$ can be defined leading to different types of function $V(I)$.  For example, suppose that $z\in Z$ partitions $X$ into a finite number of subsets.  This corresponds to a mapping $z:X\to Z$ with a constraint on the cardinality of its image $|Z|\leq e^I<|X|$.  Then, given such a partition $z:X\to Z$, one can find optimal $y(z)$ maximizing the conditional expected utility $\bE_{P(x\mid z)}\{u(x,y)\mid z\}$ for each subset $f^{-1}(z)\ni x$.  The optimal value $U(I)$ is then defined by repeating the above and optimizing over all partitions $z(x)$ satisfying the cardinality constraint $\ln|Z|\leq I$:
\begin{equation}
  U(I):=\sup_{z(x)}\left[\bE_{P(z)}\left\{\sup_{y(z)}\bE_{P(x\mid z)}\{u(x,y)\mid z\}\right\}:\ln|Z|\leq I\right]
  \label{eq:voi-hartley}
\end{equation}
Here $P(z)=P\{x\in f^{-1}(z)\}$.  The quantity $I=\ln|Z|$ is called \emph{Hartley's information}, and the difference $V(I)=U(I)-U(0)$ in this case is the value of Hartley's information.

\begin{example}
  Let $X\equiv \bR^n$ and $u(x,y)=-\frac{1}{2}\|x-y\|^2$.  Then the optimal estimator is the expected value $y=\bE\{x\}$, which is found from the stationary condition:
  \[
    \frac{\partial}{\partial y}\bE_{P(x)}\left\{-\frac{1}{2}\|x-y\|^2\right\}=y-\bE\{x\}=0
  \]
  The optimal value is $U(0)=-\frac{1}{2}\sigma^2_x$, where $\sigma^2_x$ is the variance of $x$.  Given a partition $z:X\to Z$ of $X$ into $k=|Z|$ subsets, one can compute $k$ estimators given by conditional expectations $y(z)=\bE\{x\mid z\}$.  The value $U(\ln k)$ can be estimated by computing and minimizing the average of conditional variances $\sigma^2_x(z)$ over several partitions.
  \label{ex:k-means}
\end{example}

One can see from equation~(\ref{eq:voi-hartley}) that the computation of the value of Hartley's information is quite demanding, and Example~\ref{ex:k-means} suggests that it might involve a procedure such as the $k$-means clustering algorithm or training a multilayer neural network.  Indeed, computing the error at the output layer of a perceptron and adjusting the output weights corresponds to finding optimal output function $y(z)$ in equation~(\ref{eq:voi-hartley}); back-propagation of the error into hidden layers and adjusting their weights corresponds to finding optimal partition $z(x)$ in (\ref{eq:voi-hartley}).  Although there exist efficient algorithms for such optimization, it is clear that using the value of Hartley's information is not practical due to high cost of the computations involved.  The main result of the theory \cite{Stratonovich20:_inf} is that the value of Hartley's information (\ref{eq:voi-hartley}) is asymptotically equivalent to the value of Shannon's information, which is much easier to compute.

Recall the definition of Shannon's mutual information \cite{Shannon48}:
\begin{align*}
  I(X,Y):=\bE_{W(x,y)}\left\{\ln\frac{P(x\mid y)}{P(x)}\right\}&=H(X)-H(X\mid Y)\\
  &=H(Y)-H(Y\mid X)
\end{align*}
where $W(x,y)=P(x\mid y)Q(y)$ is the joint probability distribution on $X\times Y$, and $H(\cdot)=-\bE_P\{\ln P(\cdot)\}$ is the entropy function.  The following inequality is valid:
\[
  0\leq I(X,Y)\leq\min\{H(X),H(Y)\}\leq\min\{\ln|X|,\ln|Y|\}
\]
The value of Shannon's information is defined using the quantity:
\begin{equation}
  U(I):=\sup_{P(y\mid x)}[\bE_W\{u(x,y)\}:I(X,Y)\leq I]
  \label{eq:voi-shannon}
\end{equation}
where optimization is over all conditional probabilities $P(y\mid x)$ (or joint measures $W(x,y)=P(y\mid x)P(x)$) satisfying the information constraint $I(X,Y)\leq I$.  Contrast this with $U(I)$ for Hartley's information (\ref{eq:voi-hartley}), where optimization is over the mappings $y(x)=y\circ z(x)$.  As was pointed out in \cite{Belavkin18:_igaia}, the relation between functions (\ref{eq:voi-hartley}) and (\ref{eq:voi-shannon}) is similar to that between optimal transport problems in the Monge and Kantorovich formulations.

Function $U(I)$ defined in (\ref{eq:voi-shannon}) is strictly increasing and concave, and it has the following inverse:
\begin{equation}
  I(U):=\inf[I(X,Y):\bE_W\{u(x,y)\}\geq U]
  \label{eq:iov-shannon}
\end{equation}
It is a proper convex and strictly increasing function, where it is finite.  The strictly increasing and concave (resp. convex) properties of $U(I)$ (resp. $I(U)$) can be shown in more general settings, when information is defined by any closed functional (see Proposition~3 in \cite{Belavkin11:_optim}).  This means that solutions to these conditional extremum problems can be found by the standard method of Lagrange multipliers (see \cite{Stratonovich20:_inf,Belavkin11:_optim} for details).  Thus, the optimal joint distributions belong to the following exponential family:
\begin{equation}
  W(x,y;\beta)=P(x)Q(y)e^{\beta u(x,y)-\gamma(x;\beta)} \label{eq:exponential}
\end{equation}
where $P$ and $Q$ are the marginal distributions of $W$, and function $\gamma(x;\beta)$ is defined by the normalization condition $\int_{X\times Y}dW(x,y;\beta)=1$.  Parameter $\beta$ is called the \emph{inverse temperature}, and it is the Lagrange multiplier associated to the constraint $\bE\{u\}\geq U$ in~(\ref{eq:iov-shannon}).  The temperature $\beta^{-1}$ is associated respectively to the constraint $I(X,Y)\leq I$ in~(\ref{eq:voi-shannon}).  Their values are defined by the following conditions:
\[
  \beta^{-1}=U'(I)\,,\qquad \beta=I'(U)
\]
In fact, this can also be seen from the following considerations.  Function $U(I)$ is a proper concave function, and therefore it is the Legendre-Fenchel dual (see \cite{Rockafellar74,Tikhomirov90:_convex}) of some proper concave function $F(\beta^{-1})$:
\[
  U(I)=\inf\{\beta^{-1} I-F(\beta^{-1})\}\quad\iff\quad I=F'(\beta^{-1})\quad\iff\quad \beta^{-1}=U'(I)
\]
Function $I(U)$ is a proper convex function, and therefore it is the Legendre-Fenchel dual of some proper convex function $\Gamma(\beta)$:
\[
  I(U)=\sup\{\beta U-\Gamma(\beta)\}\quad\iff\quad U=\Gamma'(\beta)\quad\iff\quad \beta=I'(U)
\]
Convex function $\Gamma(\beta)$ is the cumulant generating function of distribution~(\ref{eq:exponential}).  In particular, $U(\beta)=\Gamma'(\beta)$ is the expected value $\bE_{W(\beta)}\{u(x,y)\}$.  Concave function $F(\beta^{-1})$ is sometimes referred to as \emph{free energy}, and $I(\beta^{-1})=F'(\beta^{-1})$ is equal to Shannon's mutual information $\bE_{W(\beta)}\{\ln W-\ln (P\otimes Q)\}$ of distribution~(\ref{eq:exponential}).  Functions $F$ and $\Gamma$ have the following relation:
\[
  F(\beta^{-1})=-\beta^{-1}\Gamma(\beta)
\]

The following procedure can be used to obtain the dependencies $U(I)$ or $I(U)$ and the value of Shannon's information $V(I)=U(I)-U(0)$.   Optimal solution~(\ref{eq:exponential}) is used to define the expression for function $\Gamma(\beta)$, which is then used to derive two functions:
\[
  U(\beta)=\Gamma'(\beta)\,,\qquad I(\beta)=\beta\,\Gamma'(\beta)-\Gamma(\beta)
\]
The dependency $U(I)$ (or $I(U)$) is then obtained either parametrically from $U(\beta)$ and $I(\beta)$ or explicitly by excluding $\beta$ from one of the equations.  Alternatively, one can use free energy $F(\beta^{-1})$ and define $U(I)$ from $I(\beta^{-1})=F'(\beta^{-1})$ and $U(\beta^{-1})=\beta^{-1}I(\beta^{-1})-F(\beta^{-1})$.

Let us now consider function $\Gamma(\beta)$ for distribution~(\ref{eq:exponential}).  Taking partial traces of solution~(\ref{eq:exponential}) and using the law of total probability leads to the following system of integral equations:
\begin{align}
 \int_X dW(x,y)&=dQ(y) &\implies& &\int_X e^{\beta\,u(x,y)-\gamma(x;\beta)}\,dP(x)=1 \label{eq:intx-1}\\
 \int_Y dW(x,y)&=dP(x) &\implies& &\int_Y e^{\beta\,u(x,y)}\,dP(y)=e^{\gamma(x;\beta)} \label{eq:inty-gamma}
\end{align}
If the linear transformation $T(\cdot)=\int_X e^{\beta\,u(x,y)}(\cdot)$ has inverse, then from~(\ref{eq:intx-1}) we have $e^{-\gamma(x;\beta)}dP(x)=T^{-1}(1)$ or
\[
  \gamma(x;\beta)=-\ln\int_Y b(x,y)\,dy+\ln[dP(x)/dx]=\gamma_0(x;\beta)-h(x)
\]
where $b(x,y)$ is the kernel of the inverse linear transformation $T^{-1}$, $\gamma_0(x;\beta):=-\ln\int_Y b(x,y)\,dy$, and $h(x)=-\ln [dP(x)/dx]$ is random entropy or \emph{surprise}.  Integrating the above with respect to measure $P(x)$ we obtain
\[
  \Gamma(\beta):=\int_X\gamma(x;\beta)\,dP(x)=\Gamma_0(\beta)-H(X)
\]
where $\Gamma_0(\beta):=\int_X\gamma_0(x;\beta)\,dP(x)$.  Notice that $\Gamma'(\beta)=\Gamma'_0(\beta)=U(\beta)$, and therefore
\[
  I(\beta)=\beta\,\Gamma'(\beta)-\Gamma(\beta)=H(X)-[\Gamma_0(\beta)-\beta\,\Gamma'_0(\beta)]
\]
Function $\Gamma_0(\beta)-\beta\,\Gamma'_0(\beta)$ is clearly the conditional entropy $H(X\mid Y)$, because $I(X,Y)=H(X)-H(X\mid Y)$.

Further analysis is complicated by the dependency of solution~(\ref{eq:exponential}) on marginal distribution $P(x)$.  Generally, $P(x)$ influences not only the output distribution $Q(y)$ (i.e. as $dP(x)\mapsto \int_X dP(y\mid x)\,dP(x)=dQ(y)$), but also the conditional probability $P(x\mid y)=P(x)e^{\beta\,u(x,y)-\gamma(x;\beta)}$.  However, as was shown in \cite{Stratonovich20:_inf}, this dependency on $P(x)$ disappears, if the product $e^{-\gamma(x;\beta)}P(x)$ is independent of $x$.  Indeed, let $e^{-\Gamma_0(\beta)}=e^{-\gamma(x;\beta)}\,dP(x)/dx=\mathrm{const}$.  Then from equation~(\ref{eq:intx-1}) we obtain
\[
  e^{-\Gamma_0(\beta)}\int_X e^{\beta\,u(x,y)}dx=1\quad\implies\quad \Gamma_0(\beta)=\ln\int_X e^{\beta\,u(x,y)}\,dx
\]

It turns out that $e^{-\gamma(x;\beta)}\,dP(x)/dx=\mathrm{const}$, if the objective function is translation invariant: $u(x,y)=u(x+z,y+z)$.  Indeed, using translation invariance and equation~(\ref{eq:intx-1}) gives
\[
  \int_X e^{\beta\,u(x+z,y+z)-\gamma(x+z;\beta)}\,dP(x+z)=\int_X e^{\beta\,u(x,y)-\gamma(x+z;\beta)}\,dP(x+z)=1
\]
Combining this with equation~(\ref{eq:intx-1}) implies that
\[
  e^{-\gamma(x+z;\beta)}dP(x+z)/dx=e^{-\gamma(x;\beta)}dP(x)/dx=\mathrm{const}
\]
Many objective functions $u(x,y)$ are defined using the difference $x-y$, which means they are translation invariant.

\begin{example}[Squared error and Gaussian case]
  Let $u(x,y)=-\frac12(x-y)^2$.  Then $u(x,y)=u(x+z,y+z)$, and
  \begin{align*}
    \Gamma_0(\beta)&=\ln\int_{-\infty}^\infty e^{-\frac12\,\beta\,(x-y)^2}\,dx=\ln\sqrt{\frac{2\pi}{\beta}}\\
    U(\beta)&=\Gamma_0'(\beta)=-\frac{1}{2\beta}\\
    I(\beta)&=-\frac{1}{2}-\Gamma(\beta)=-\frac{1}{2}+H(X)-\Gamma_0(\beta)=H(X)-\frac{1}{2}\left[\ln(2\pi)+1-\ln\beta\right]
  \end{align*}
  The latter expression allows us to express $\beta=2\pi e^{2[I-H(X)]+1}$ and write explicit dependency
  \begin{equation}
    U(I)=-\frac{1}{4\pi} e^{2[H(X)-I]-1} \label{eq:ui-quadratic}
  \end{equation}
  The value of information in this case is
  \[
    V(I)=U(I)-U(0)=\frac{1}{4\pi} e^{2H(X)-1}\left(1-e^{-2I}\right)
  \]
  For Gaussian density $dP(x)/dx=\frac{1}{\sqrt{2\pi\sigma^2_x}}\,e^{-\frac{x^2}{2\sigma^2_x}}$ we have
  \[
    H(X)=\frac{1}{2}\left[\ln(2\pi\sigma^2_x)+1\right]\,,\qquad e^{2H(X)-1}=2\pi\sigma^2_x
  \]
  and in this case
  \[
    U(I)=-\frac{1}{2}\sigma^2_x e^{-2I}\,,\qquad V(I)=\frac{1}{2}\sigma^2_x(1-e^{-2I})
  \]
  \label{ex:quadratic}
\end{example}

\begin{example}[Root-mean-square error]
  The root-mean-square error (RMSE or standard error) is one of the most important criteria to evaluate data-driven models.  The result from Example~\ref{ex:quadratic} can be used to compute the smallest RMSE as a function of information.  Indeed, $\mathrm{RMSE}(I)=\sqrt{-2U(I)}$, where $U(I)$ is given by equation~(\ref{eq:ui-quadratic}):
  \[
    \mathrm{RMSE}(I)=\frac{1}{\sqrt{2\pi e}}\,e^{H(X)-I}
  \]
  If $x$ is assumed to have normal distribution with variance $\sigma^2_x$, then $e^{H(X)}=\sigma_x\sqrt{2\pi e}$ and
  \begin{equation}
    \mathrm{RMSE}(I)=\sigma_x\,e^{-I} \label{eq:rmse-i}
  \end{equation}
  If the amount of information $I$ can be estimated from data (e.g. as mutual information $I(X,Z)$ between the predictors and response variables), then the functions above define the smallest possible standard error.
\end{example}

\section{Application: Analysis of forecasts of cryptocurrency log-returns}\label{sec:rmse}

In this section, we illustrate how the value of information can facilitate the analysis of performance of data-driven models.  Here we use time-series forecasts applied to daily log-returns of cryptocurrency exchange rates.

\begin{figure}
\centering
\includegraphics[width=.49\textwidth]{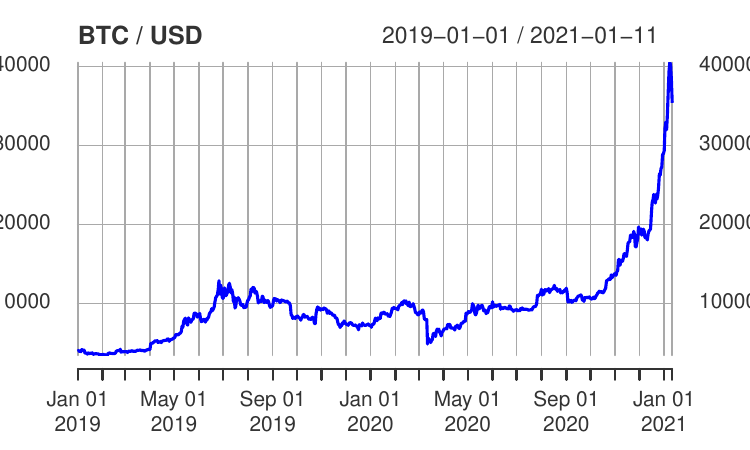}
\includegraphics[width=.49\textwidth]{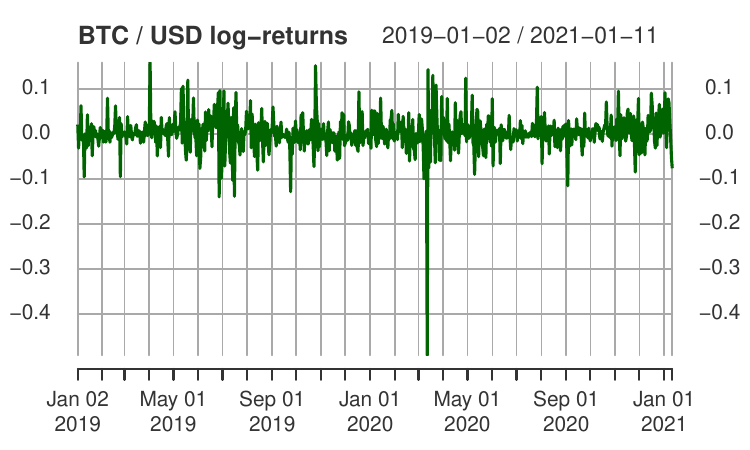}
\caption{Close day prices of BTC/USD (left) and the corresponding log-returns (right).}\label{fig:btcusd}
\end{figure}

\begin{figure}
\centering
\includegraphics[height=4.5cm]{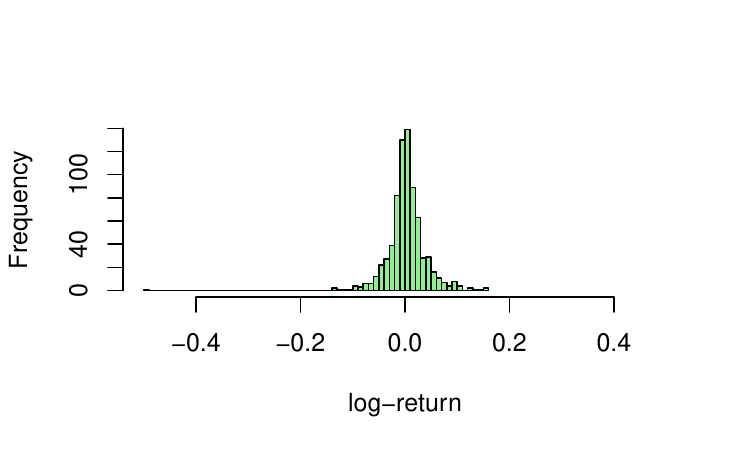}
\includegraphics[height=4.5cm]{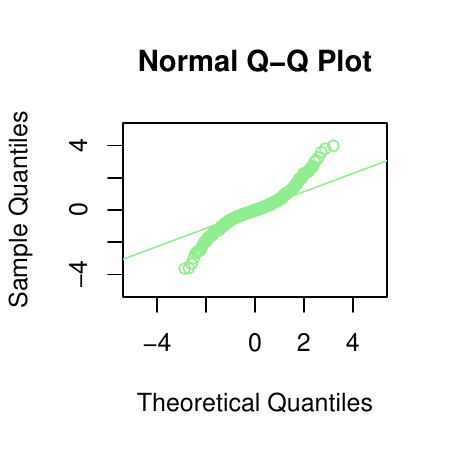}
\caption{Distribution of BTC/USD log-returns (left) and its comparison with normal distribution (right).}\label{fig:hist}
\end{figure}

The dataset used contains daily prices $s(t)$ of several cryptocurrency pairs during the period between Jan 1, 2019 and Jan 11, 2021.  Figure~\ref{fig:btcusd} shows an example of prices of Bitcoin in US Dollars (BTC/USD) and the corresponding log-returns, which are defined as
\[
  r(t+1):=\ln\left[\frac{s(t+1)}{s(t)}\right]
\]
Figure~\ref{fig:hist} shows the distribution of log-returns $r(t)$ for BTC/USD.  They are approximately zero-mean with $r(t)>0$ corresponding to a price increase and vice versa.  Although it is quite common to model log-returns by a Gaussian distribution, it is easy to see that the distribution has heavy tails (see the QQ-plot on Figure~\ref{fig:hist} comparing the distribution with a Gaussian), and some extreme price changes are not unusual (e.g. notice the significant price decrease on March~12, 2020, which was caused by the announcements related to the COVID-19 pandemic).

\begin{figure}
\centering
\includegraphics[height=4.5cm]{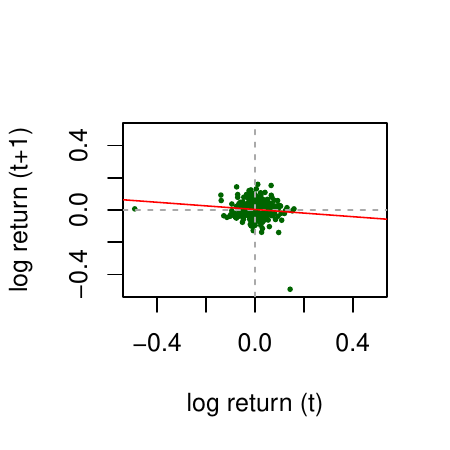}
\includegraphics[height=4.5cm]{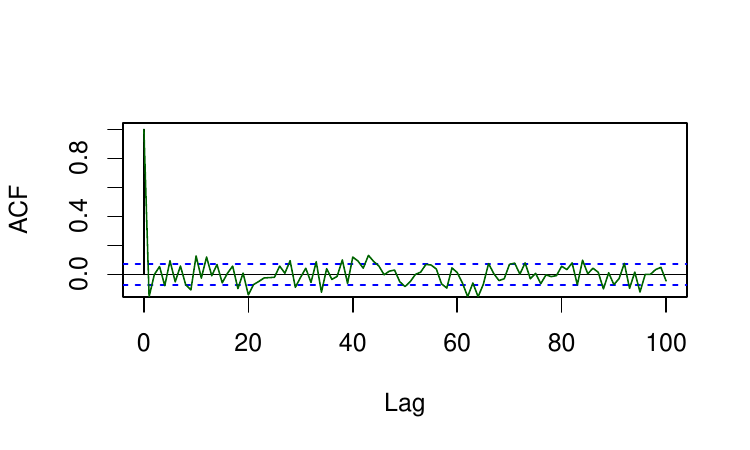}
\caption{Relation between log-returns on two consecutive days (left) and the autocorrelation function (right).}\label{fig:acf}
\end{figure}

Predicting price changes is very challenging.  In fact, the existence of such forecasts would create an arbitrage, which should quickly disappear in an open market.  The left chart on Figure~\ref{fig:acf} plots log-returns for two consecutive days: $r(t)$ (abscissa) and $r(t+1)$ (ordinates).  One can see that there is no obvious relation between $r(t)$ and $r(t+1)$, and they are often assumed to be independent (and hence prices $s(t)$ are often modelled by a Markov process).

On the other hand, in continuous time independence of log-returns would mean that $\{r(t)\}$ is a so-called $\delta$-correlated stochastic process (i.e. its autocorrelation function is proportional to the Dirac $\delta$-function).  It is well-known that such processes are unphysical, because any $\delta$-correlated stochastic process must have infinite variance $\sigma^2$ (indeed, one can show that $\sigma^2$ is the integral of spectral density, which is the Fourier transform of the autocorrelation function; the Fourier image of the $\delta$-function is a constant function \cite{Stratonovich:_noise}).  Therefore, there must be some small information about future log-return $r(t+1)$ contained in the past values $r(t),r(t-1),\ldots,r(t-n)$.  This can be seen from the plot of the autocorrelation function for BTC/USD shown on the right chart of Figure~\ref{fig:acf}.

The idea of autoregressive models is to use the small amount of information between the past and future values for forecasts.  Here, we shall employ several techniques to learn models $y=f(z)$, where the predictor $z=(r(t),r(t-1),\ldots,r(t-n))$ is a vector of previous values of log-returns, and the model output $y(z)$ is the forecast of the unknown future log-return $x=r(t+1)$ (the response).  The hypothesis is that increasing the number $n$ of lags should increase the amount of information used for the forecasts.

In addition to autocorrelations (correlations between the values of $\{r(t)\}$ at different times), information can be increased by using cross-correlations (correlations between log-returns of different symbols in the dataset).  Thus, the vector of predictors is an $m\times n$-tuple, where $m$ is the number of symbols used, and $n$ is the number of time lags.  In this paper we report result of predicting log-returns of BTC/USD using the range $m\in\{1,2,\ldots,5\}$ of symbols (BTC/USD, ETH/USD, DAI/BTC, XRP/BTC, IOT/BTC) and $n\in\{2,3,\ldots,20\}$ of lags.  This means that the models used predictors $(z_1,\ldots,z_{m\times n})$, where $m\times n$ ranged from 2 to 100.

\begin{figure}
\centering
\includegraphics[width=.49\textwidth]{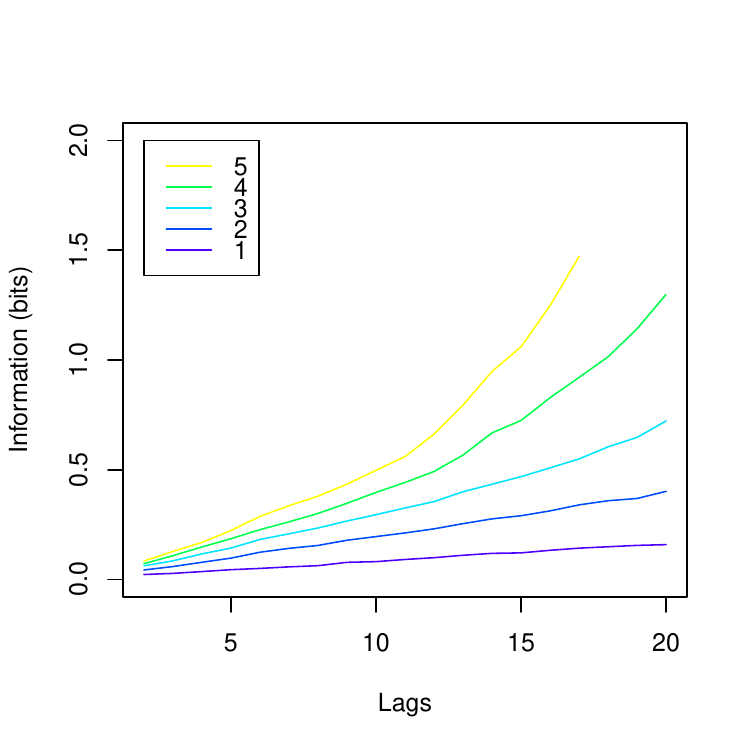}
\includegraphics[width=.49\textwidth]{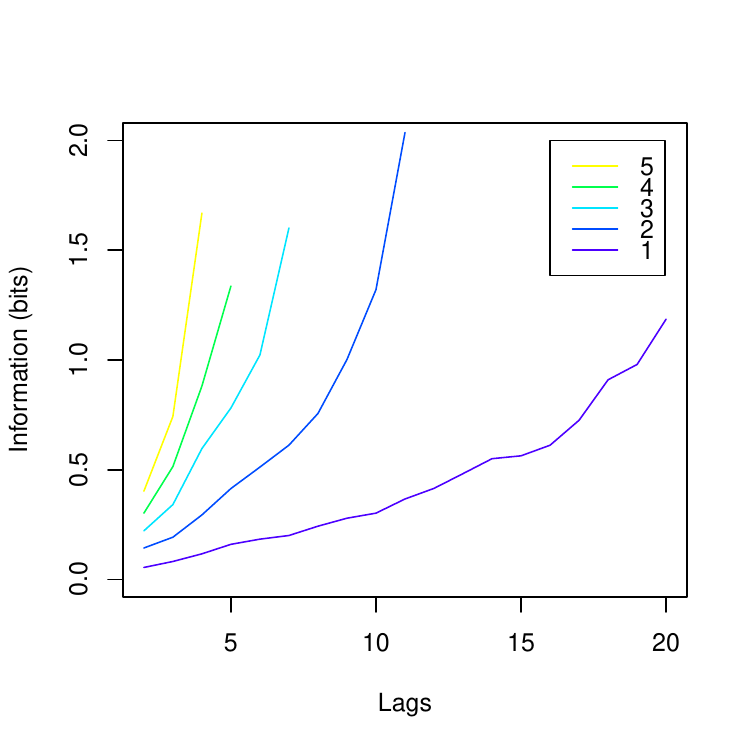}
\caption{The average amount of mutual information between predictors and response in the training sets (left) and test sets (right).  Abscissa shows the number $n$ of lags, and different curves correspond to different numbers $m$ of symbols used.}\label{fig:information}
\end{figure}

In order to analyse the performance of models using the value of information, one has to estimate the amount of information between the predictors $z_1,\ldots,z_{m\times n}$ and the response variable $x$.  Here we employ the following Gaussian formula for Shannon's mutual information \cite{Stratonovich20:_inf}:
\[
  I(X,Z)\approx\frac{1}{2}\left[\ln\det K_z + \ln\det K_x - \ln\det K_{z\oplus x}\right]
\]
where $K_z$ is the covariance matrix of predictors $z\in\bR^{m\times n}$, $K_x$ is the covariance of response $x$ (for one dimension $\det K_x=\sigma^2_x$), and $K_{Z\oplus X}$ is the covariance of $Z\oplus X$.  We use the approximate sign $\approx$, because the distributions of log-returns are generally not Gaussian (in fact, the above formula gives a lower bound for non-Gaussian random variables).  Natural logarithm corresponds to measuring information in `nats'; for `bits' one has to use $\log_2$.

For each collection of predictors $(z_1,\ldots,z_{m\times n})$ and response $x$, the data was split into multiple training and testing subsets using the following rolling window procedure.  Here we used 100 and 25 days data windows for training and testing respectively.  After training and testing the models, the windows were moved forward by 25 days.  Thus, the data of approximately 700 days (Jan 2019 to Jan 2021) was split into $(700 - 100) / 25 = 24$ pairs of training and testing sets.  The results reported here are the average results from these 24 subsets.

Figure~\ref{fig:information} shows the average amounts of information $I(X,Z)$ in the training sets (left) and testing sets (right).  Information (ordinates) is plotted against the number $n$ of lags (abscissa) and for $m\in\{1,2,\ldots,5\}$ symbols (different curves).  The data was used to train and test the following types of models:
\begin{enumerate}
\item Multiple mean-square linear regression (LM).
\item Partial least squares regression (PLS).
\item Feed-forward neural network (NN).
\end{enumerate}
The first model has no hyperparameters; the PLS regression used here employed SIMPLS algorithm \cite{DeJong1993} with 3 components; NN used here had just one hidden layer with 3 logistic units and trained for 30 epochs.  This is admitably not an optimal choice of models, but finding the best model or a set of hyperparameters was not the purpose of this study.  The models were used to illustrate their performance from the point of the value of information theory.

\begin{figure}
\centering
\includegraphics[width=.32\textwidth]{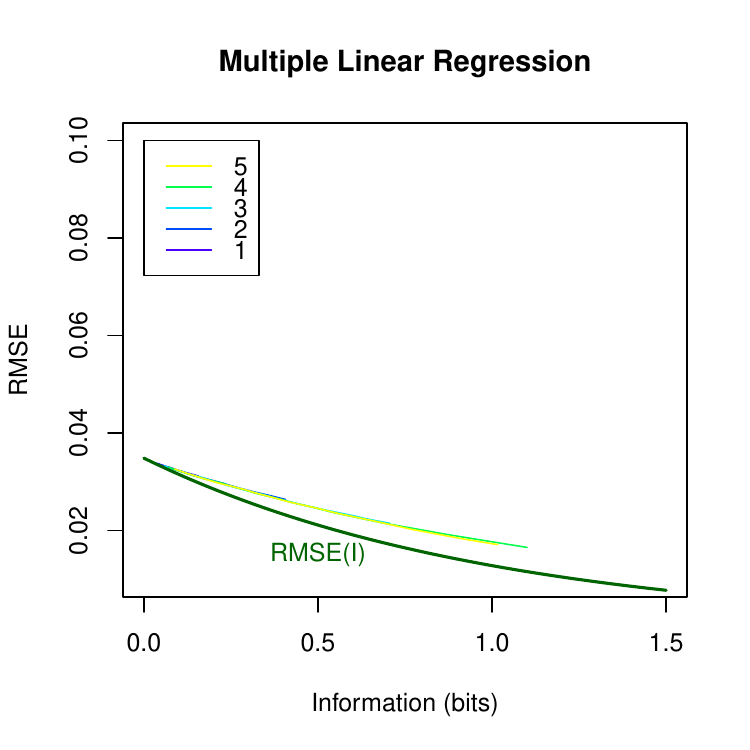}
\includegraphics[width=.32\textwidth]{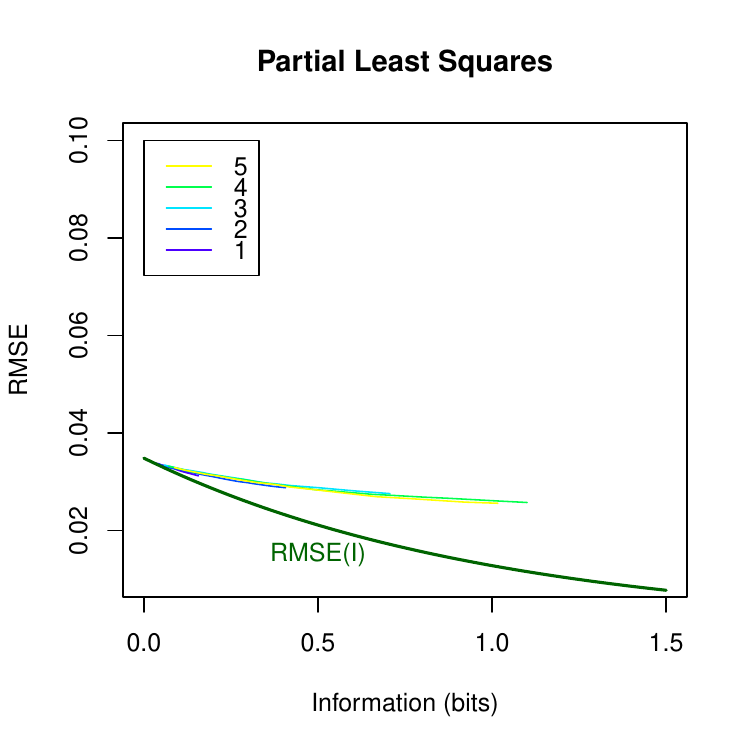}
\includegraphics[width=.32\textwidth]{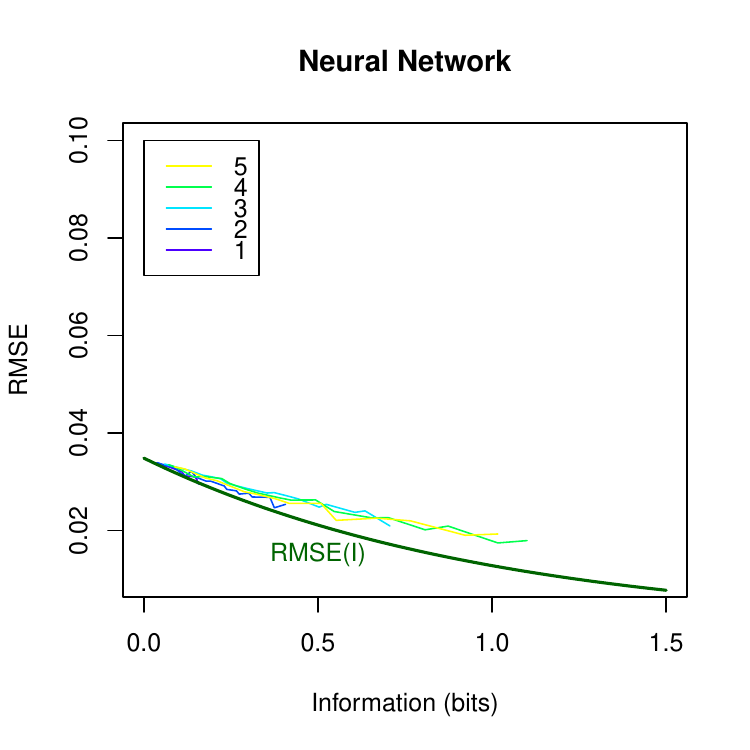}
\caption{RMSE results of fitted values of three types of models on training data as functions of information in the training data.  Theoretical $\mathrm{RMSE}(I)$ curve~(\ref{eq:rmse-i}) is plotted for standard deviation of response $\sigma_x\approx .0386$ estimated from the training sets.}\label{fig:rmse-train}
\end{figure}

\begin{figure}
\centering
\includegraphics[width=.32\textwidth]{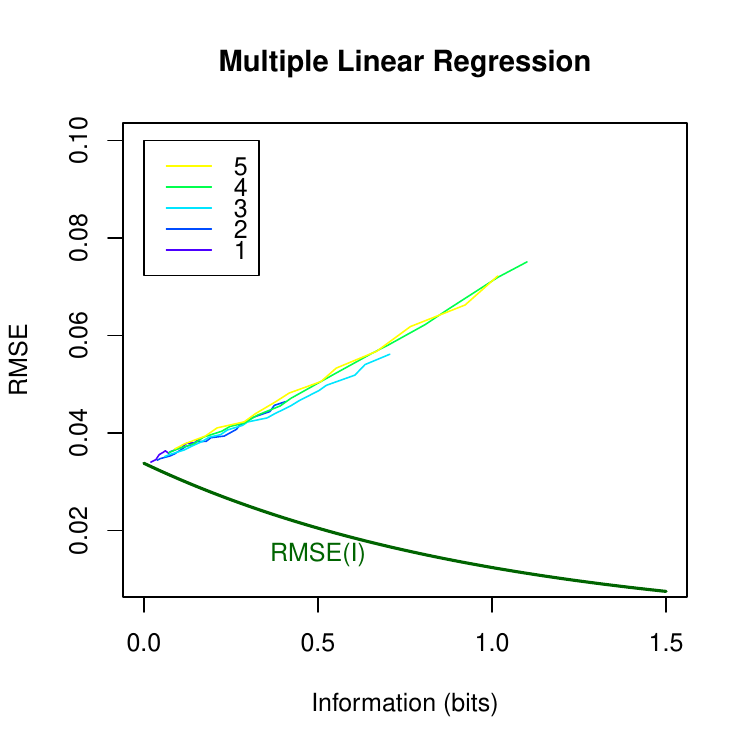}
\includegraphics[width=.32\textwidth]{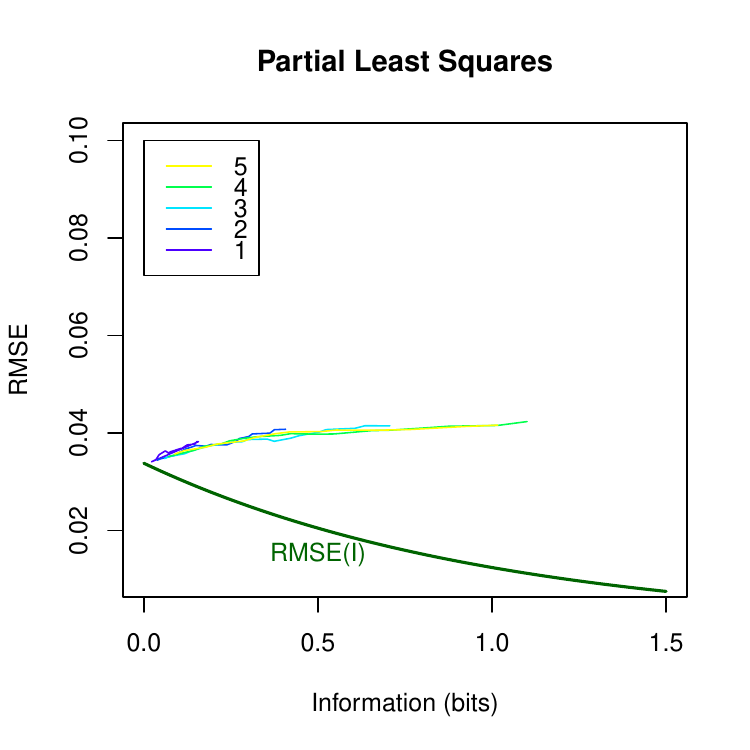}
\includegraphics[width=.32\textwidth]{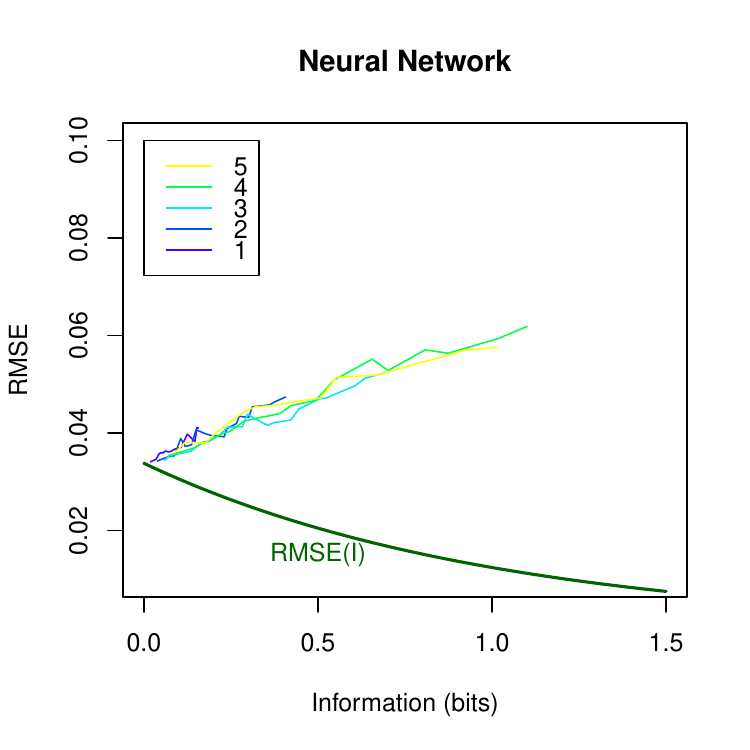}
\caption{RMSE results of predicted values from three types of models on testing data as functions of information in the training data.  Theoretical $\mathrm{RMSE}(I)$ curve~(\ref{eq:rmse-i}) is plotted for standard deviation of response $\sigma_x\approx .0361$ estimated from the testing sets.}\label{fig:rmse-test}
\end{figure}

Figures~\ref{fig:rmse-train} and \ref{fig:rmse-test} show standard errors (RMSE) of the models as function of the information amount $I$ contained in the training data.  Different curves are plotted for different numbers of symbols $m\in\{1,\ldots,5\}$.  Theoretical lower bounds are shown by the $\mathrm{RMSE}(I)$ curves computed using formula~(\ref{eq:rmse-i}) with standard devition of response $x$ estimated from the training and testing sets.  Figure~\ref{fig:rmse-train} shows RMSE of the models fitting the training data after training, while Figure~\ref{fig:rmse-test} shows the errors of prediction on testing data.  The following observations can be made from the results shown on Figures~\ref{fig:rmse-train} and \ref{fig:rmse-test}:
\begin{enumerate}
\item Errors of fitting the training data closely follow theoretical curve $\mathrm{RMSE}(I)$.  One can see that LM and NN achieve errors on the training data close to theoretical.  PLS has higher errors, which can be explained by the fact that the aim of the PLS algorithm is not to minimize squared errors, but to maximize covariance between predictors and reponse \cite{DeJong1993}.
\item All models show higher errors on the testing data.  PLS achieved smaller and more stable errors in forecasts than LM or NN in this experiment.
\item Increasing information leads to decreasing errors on the training data, but not necessarily on new data (testing or prediction).
\item Models using $m>1$ symbols achieve smaller errors on the testing data than models with just one symbol.  We note also that when using $m=4$ or $5$ symbols, the amount of information of say $I=.1$~bits can be achieved using only $n\leq 5$ lags (see left chart on Figure~\ref{fig:information}).  The same amount of information in data with $m=1$ symbol requires $n>20$ lags.  Thus, cross-correlations potentially provide more valuable information for forecasts than autocorrelations.
\item Linear models, and in particular PLS, appear to have more robust performance than the simple neural network used here.  The large variance of standard errors for NN shown on Figures~\ref{fig:rmse-train} and \ref{fig:rmse-test} are potentially due to random initialization and higher uncertainty in the setting of hyper-parameters (e.g. hidden nodes, the number of epochs to train, activation functions).
\end{enumerate}

\begin{remark}
  RMSE can also be plotted against mutual information in the test set shown on the right chart of Figure~\ref{fig:information}.  However, this information was not used to learn the models, and hence we do not report these plots here.  One can also notice from Figure~\ref{fig:information} that mutual information in the test sets achieves higher values (approaching 2 bits) than in the training sets.  This can be explained by random effects, as the test sets were four times smaller than the training sets.
\end{remark}

\begin{figure}
\centering
\includegraphics[width=.32\textwidth]{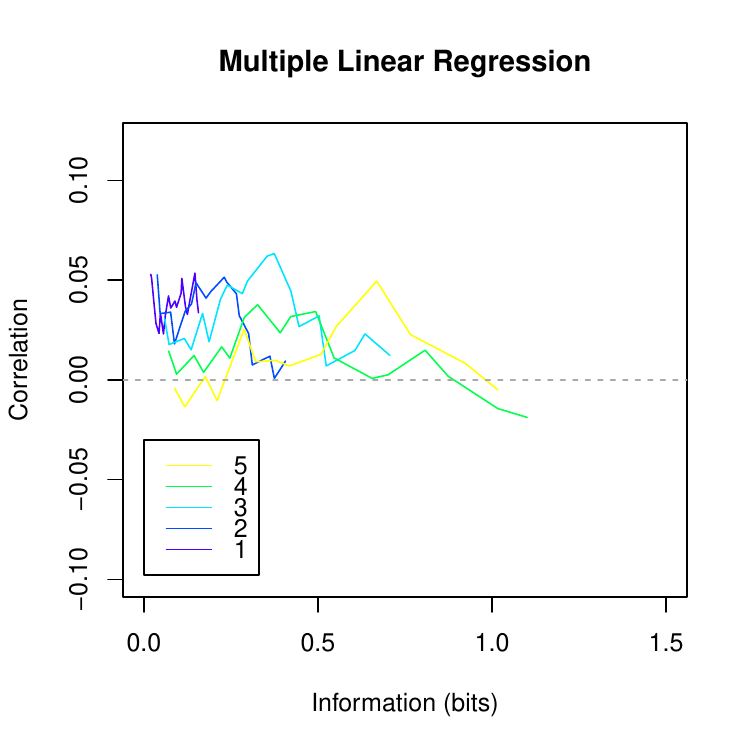}
\includegraphics[width=.32\textwidth]{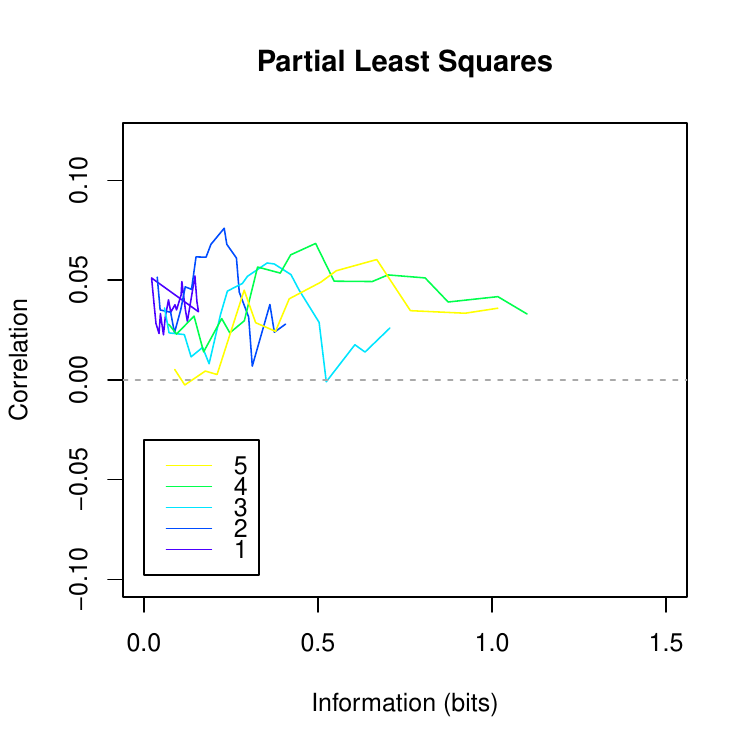}
\includegraphics[width=.32\textwidth]{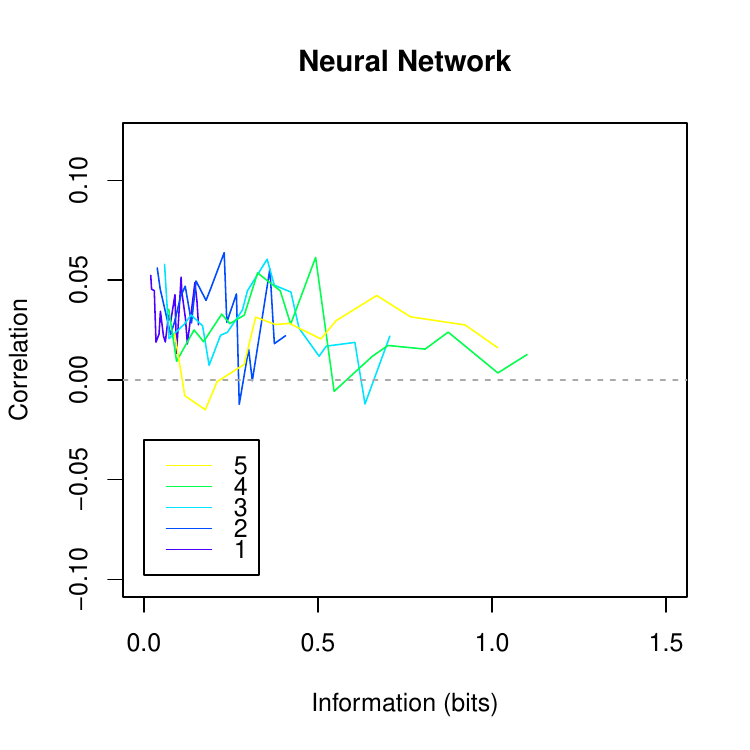}
\caption{Correlation between predicted values from models and desired response in the test data as functions of information in the training sets.}\label{fig:cor}
\end{figure}

Let us point out that RMSE is a general, but certainly not the only and potentially not the most useful measure to assess model's performance.  Figure~\ref{fig:cor} reports correlations between the predicted and the desired log-returns (i.e. correlation between the model output $y(z)$ and the desired response $x$).  One may notice that the best linear models (LM and PLS) are those using $m\in\{2,3\}$ symbols, and the maximum correlations are generally achieved at higher amounts of information than those achieving the minimums of RMSE.

\begin{figure}
\centering
\includegraphics[width=.32\textwidth]{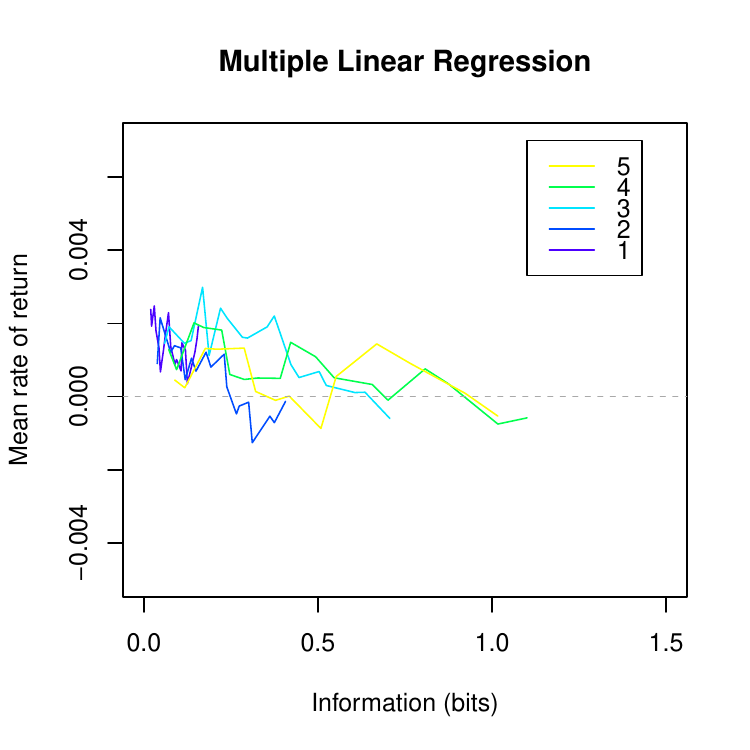}
\includegraphics[width=.32\textwidth]{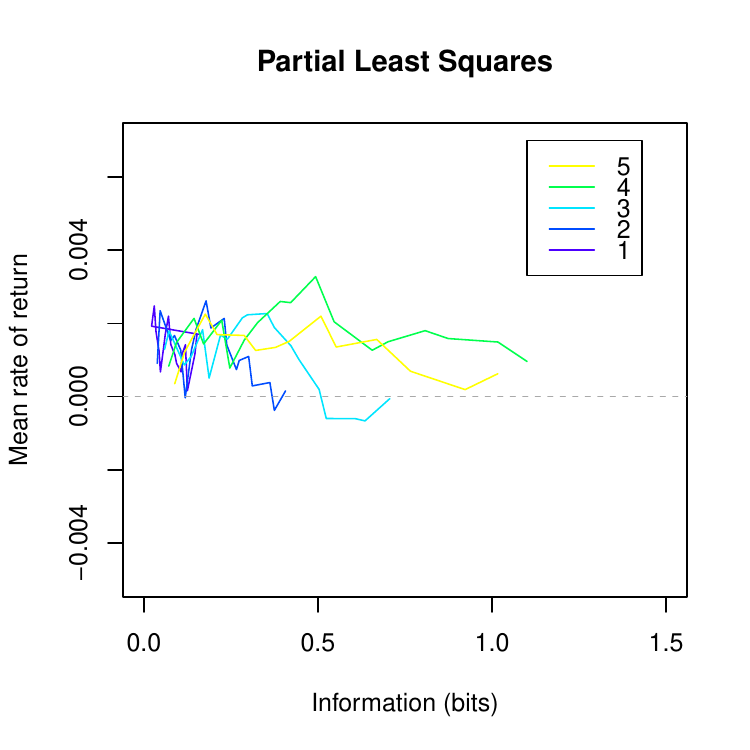}
\includegraphics[width=.32\textwidth]{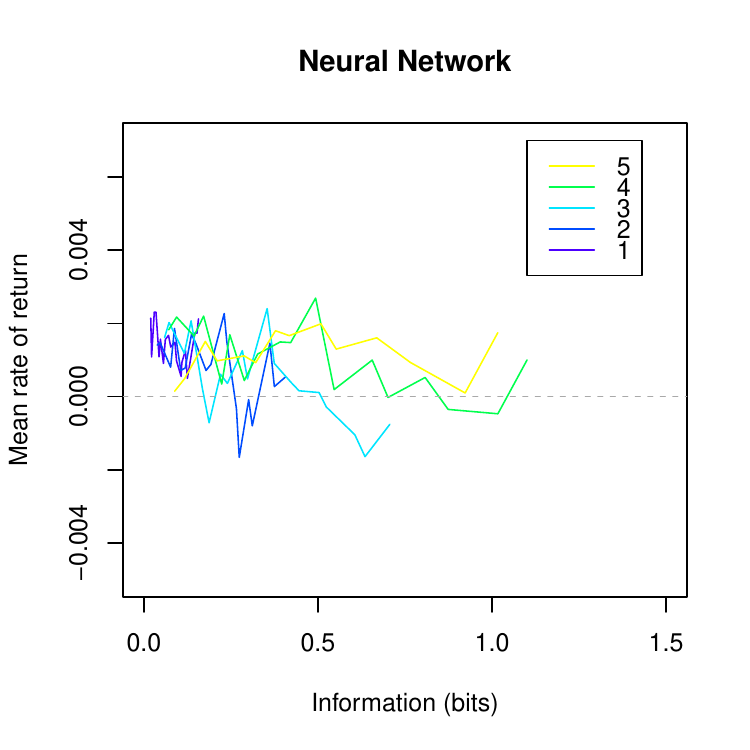}
\caption{Mean rates of return as functions of information for different models.}\label{fig:mrr}
\end{figure}

Finally, we estimated the mean rates of return (MRR) from the model forecasts, if they were used for trading.  Here, we used the following formula:
\[
  \mathrm{MRR}:=e^{\bE\{\operatorname{sign}(y(z))\operatorname{sign}(x)|x|\}}-1
\]
where $y(z)$ is the predicted log-return, $x$ is the `true' log-return from the test data, and $\operatorname{sign}$ is the signum function.  Thus, when the signs of $y(z)$ and $x$ coincide, then the log-return from trading is positive $|x|$; otherwise, the log-return is $-|x|$.  The expected value $\bE\{\operatorname{sign}(y(z))\operatorname{sign}(x)|x|\}$ is the mean log-return from trading $\langle r\rangle$, which is converted into the effective rate of return by the formula $e^{\langle r\rangle}-1$.  Thus, the value of $\mathrm{MRR}=.01$ means 1\% return per day without taking into account trading fees.  Figure~\ref{fig:mrr} reports the estimated mean rates of return for the three types of models.  Some models achieve mean rates of return $.3\%$ and $.4\%$ per day, which is slightly higher than the average rate of return of $.26\%$ from BTC / USD in the testing sets.  Note also that the mean rate of return from the models can also be as low as $-.5\%$ per day.

\section{Discussion}\label{sec12}

We have reviewed the main mathematical ideas of the value of information theory in the context of translation invariant objective functions.  These functions are important for data-driven models, such as the mean-square cost or standard error.  We have derived simple expressions for the lower bound of RMSE as a function of mutual information and applied it to the analysis of performance of time-series forecasts using cryptocurrency data.  We showed how these information-theoretic ideas can enrich our understanding of data and the models and potentially lead to a more intelligent learning and optimization of model parameters.

\subsubsection{Acknowledgements}

Stefan Behringer is deeply acknowledged for additional discussion of the example, Roman Tarabrin is deeply acknowledged for providing a MacBookPro laptop used for the computational experiments.  This research was funded in part by the ONR grant number N00014-21-1-2295.

\bibliographystyle{splncs04}
\bibliography{rvb-lion22}

\end{document}